\pdfoutput=1
\documentclass[manuscript,screen,nonacm,10pt]{acmart}

\usepackage{booktabs}
\usepackage{cleveref}
\usepackage{footmisc}
\usepackage{fp}
\usepackage{listings}
\usepackage{numprint}
\usepackage{xspace}

\newcommand{\sys}{{Exoshuffle-CloudSort}\xspace}
\newcommand{\bench}{{100\,TB CloudSort Benchmark}\xspace}

\nprounddigits{0}
\newcommand{\code}[1]{\lstinline[basicstyle=\small\ttfamily]|#1|}

\begin{document}

\title{\sys}

% Tweaks to acmart template
\makeatletter
% Stop complaining about no countries in affiliations
\def\@ACM@checkaffil{}
% Do not print the author addresses footnote block
\let\@authorsaddresses\@empty
\makeatother

\newcommand{\ucberkeleyaffiliation}{
\affiliation{%
  \institution{UC Berkeley}
%   \city{Berkeley}
%   \state{CA}
%   \country{USA}
}}
\newcommand{\anyscaleaffiliation}{
\affiliation{%
  \institution{Anyscale}
%   \city{San Francisco}
%   \state{CA}
%   \country{USA}
}}

\author{Frank Sifei Luan}
\authornote{Author's address: lsf@berkeley.edu, 465 Soda Hall, Berkeley, CA, USA.}
\orcid{0001-8709-6823}
\email{lsf@berkeley.edu}
\ucberkeleyaffiliation

\author{Stephanie Wang} \ucberkeleyaffiliation \anyscaleaffiliation
\author{Samyukta Yagati} \ucberkeleyaffiliation
\author{Sean Kim} \ucberkeleyaffiliation
\author{Kenneth Lien} \ucberkeleyaffiliation
\author{Isaac Ong} \ucberkeleyaffiliation
\author{Tony Hong} \ucberkeleyaffiliation
\author{SangBin Cho} \anyscaleaffiliation
\author{Eric Liang} \anyscaleaffiliation
\author{Ion Stoica} \ucberkeleyaffiliation \anyscaleaffiliation

\renewcommand{\shortauthors}{Luan et al.}

\maketitle

\FPeval\decimals{4}

\FPeval\theoreticaljct{100 * 1000 / 40 * (1/2.9 + 1/2.2 + 1/3 + 1/3)}
\FPeval\theoreticaljctR{round(\theoreticaljct:\decimals)}

\FPeval\hoursinmonth{730}
\FPeval\gb{pow(9, 10)}
\FPeval\tb{1000 * \gb}

\FPeval\totaldatasize{100 * \tb}
\FPeval\inputpartsize{2 * \gb}
\FPeval\nummappers{trunc(\totaldatasize / \inputpartsize:0)}
\FPeval\outputpartsize{4 * \gb}
\FPeval\numreducers{trunc(\totaldatasize / \outputpartsize:0)}
\FPeval\numworkers{40}
\FPeval\numreducersperworker{trunc(\numreducers / \numworkers:0)}

% (5361 + 5348 + 5426) / 3
\FPeval\jctseconds{5378}
\FPeval\jct{round(\jctseconds / 3600 : \decimals)}
% (1852 + 1852 + 1906) / 3
\FPeval\reduceseconds{1870}
\FPeval\reducetime{round(\reduceseconds / 3600 : \decimals)}

\FPeval\masternodehourlycost{0.504}
\FPeval\workernodehourlycost{1.373}
\FPeval\ebsvolumehourlycost{round(0.08 / \hoursinmonth * 40 : \decimals)}
\FPeval\vmtotalhourlycost{round(\masternodehourlycost + \workernodehourlycost * \numworkers + \ebsvolumehourlycost * (\numworkers + 1) : \decimals)}
\FPeval\computetotalcost{round(\vmtotalhourlycost * \jct : \decimals)}

\FPeval\sssgbmonthlycost{0.0225}
\FPeval\ssshourlycost{\sssgbmonthlycost * 1000 / \hoursinmonth}
\FPeval\ssstotalhourlycost{round(\ssshourlycost * 100 : \decimals)}
\FPeval\datastorageinputcost{round(\ssstotalhourlycost * \jct : \decimals)}
\FPeval\datastorageoutputcost{round(\ssstotalhourlycost * \reducetime : \decimals)}
\FPeval\datastoragetotalcost{round(\datastorageinputcost + \datastorageoutputcost : \decimals)}

\FPeval\sssgetsize{16 * 1024 * 1024}
\FPeval\sssnumgetpertask{trunc(\inputpartsize / \sssgetsize + 1 : 0)}
\FPeval\ssstotalget{trunc(\sssnumgetpertask * \nummappers : 0)}
\FPeval\sssgetprice{0.0004 / 1000}
\FPeval\sssgettotalcost{round(\sssgetprice * \ssstotalget : \decimals)}

\FPeval\sssputsize{100 * 1000 * 1000}
\FPeval\sssnumputpertask{trunc(\outputpartsize / \sssputsize + 1 : 0)}
\FPeval\ssstotalput{trunc(\sssnumputpertask * \numreducers : 0)}
\FPeval\sssputprice{0.005 / 1000}
\FPeval\sssputtotalcost{round(\sssputprice * \ssstotalput : \decimals)}

\FPeval\dataaccesstotalcost{round(\sssgettotalcost + \sssputtotalcost : \decimals)}

\FPeval\totalcost{round(\computetotalcost + \datastoragetotalcost + \dataaccesstotalcost : \decimals)}
\FPeval\totalcostW{round(\totalcost : 0)}

\section{Introduction}
\label{sec:intro}

We present \sys, a sorting application running on top of Ray using the Exoshuffle architecture~\cite{exoshuffle}.
\sys runs on Amazon EC2, with input and output data stored on Amazon S3.
Using $\numworkers{}\times$ \textsf{i4i.4xlarge} workers,
\sys completes the \bench (Indy category~\cite{cloudsort}) in \jctseconds{} seconds, with an average total cost of \$\totalcostW{}.

\section{Implementation}
\label{sec:impl}

\subsection{Overview}
\label{sec:impl:overview}

\sys is a distributed futures program running on top of Ray, a task-based distributed execution system.
The program acts as the control plane to coordinate map and reduce tasks;
the Ray system acts as the data plane, responsible for executing tasks, transferring blocks, and recovering from failures.

\sys implements a two-stage external sort algorithm.
The first stage is map and shuffle.
Each map task reads an input partition, sorts it, and partitions the result into $W$ output partitions, each sent to a merger on a worker node.
A merger receives $W$ map output partitions, merges and sorts them, and further partition the result into $R/W$ output partitions, all of which are spilled to local disk.

The second stage is reduce.
Once the map and shuffle stage finishes, each reduce task reads $W$ shuffled partitions, merges and sorts them, and writes the final output partition.

For the \bench, we set the following parameters:
\begin{itemize}
  \item Total data size is 100\,TB.
  \item Number of input partitions $M=\numprint{\nummappers{}}$. Each input partition is 2\,GB.
  \item Number of workers $W=\numworkers{}$.
  \item Number of output partitions $R=\numprint{\numreducers{}}$.
\end{itemize}

\subsection{Preparation}

The first step in \sys is to compute the partition boundary values.
For a sort record with 10-byte key, we view the first 8 bytes as a 64-bit unsigned integer partition key.
We partition the key space $[0, 2^{64}-1)$ into $R=\numprint{\numreducers{}}$ equal ranges, such that all the records within a key range should be sent to one reducer.

Every $R_1 = R / W = \numreducersperworker{}$ reducer ranges are combined into a worker range, and records in each worker range will be sent to one worker node. This yields $W = \numworkers{}$ equally-partitioned worker ranges.

% For skewed distribution, \sys can be configured to run a sampling stage 

\subsection{Map and Shuffle Stage}

In the map and shuffle stage, \sys schedules the $M=\numprint{\nummappers{}}$ map tasks onto all worker nodes.
In our experiments we set the map parallelism, i.e. the number of map tasks running on a single worker node, to be $3/4$ of the total number of vCPU cores.
Extra tasks are queued on the driver node.
Whenever a worker node finishes a map task, the driver assigns a new task from the queue to this node.

In a map task, we first download the input partition from S3.
We then sort the input data in memory, then partition it into $W=\numworkers{}$ slices.
Each slice is eagerly sent to a merge controller on each worker.
The map task returns when all slices are sent.

On the receiving end, the merge controller accumulates the map blocks in memory until a threshold is reached.
We set the threshold to 40 blocks, or about 2\,GB of data.
Once the threshold is reached, the controller launches a merge task to merge the already-sorted map blocks, and further partitions it into $R_1=\numreducersperworker{}$ merged blocks, each corresponding to a reduce task on this node.
These blocks are spilled to the local SSD for use by the reducers.

The merge parallelism is set to be the same as the map parallelism.
When the number of merge tasks reaches the maximum parallelism, and the merge controller's in-memory buffer is filled up, it will hold off acknowledging the receipt of a map block until a merge task finishes and a new merge task can launch.
This effectively creates back pressure to the map task scheduler to ensure the map, shuffle, and merge progresses are in sync.

In our experiments, the average map task duration is 24 seconds; 15 seconds are used for downloading input data. The average shuffle time (i.e. time to send and receive blocks) is 7 seconds. The merge task takes 17 seconds on average.

\subsection{Reduce Stage}

Once all map and merge tasks finish, \sys enters the reduce stage.
Each reduce task loads $R_1=\numreducersperworker{}$ from the local SSD, merges them, and uploads the sorted output partition to S3.
In our experiments, each reduce task takes 22 seconds on average.
% Ray provides pipelining of loading from disk; application pipelines uploading to S3 with merge.

\subsection{The Execution System}

A highlight of the Exoshuffle architecture is that the application program only implements the control plane logic, and the distributed futures system, Ray, handles execution.
This is reflected in \sys.
Here is an incomplete list of features provided by Ray that we take ``for free'':
\begin{itemize}
\item Task scheduling: The program specifies when and where to schedule tasks; the system handles the RPC, serialization, and other bookkeeping.
\item Network transfer: The program instructs data to be transferred by passing distributed futures as task arguments; the system implements high-performance network transfer.
\item Memory management and disk spilling: The program manipulates data references in a virtual, infinite address space; the system uses reference counting to manage distributed memory, spills objects to local disks when memory is low, and restores objects from local disks when they are needed.
\item Pipelining of network and disk I/O: The network transfer, spilling and recovery of objects are transparent to the application and are performed asynchronously. For example, the system shuffles map output blocks while other map and merge tasks are running; it spills merge task output to disk while other merge tasks are executing, and it restores merged blocks while reduce tasks are executing.
\item Fault tolerance: this is transparent to the application: the system automatically retries the operation when it encounters network failures and worker process failures.
\end{itemize}

For more details, we refer the reader to the Ray Architecture Whitepaper~\cite{raywhitepaper}, the ownership design for distributed futures systems~\cite{ownership}, and the Exoshuffle paper~\cite{exoshuffle}.

\subsection{Source Code}

\sys is implemented in about 1000 lines of Python, and about 300 lines of C++. The C++ component implements two functionalities: sorting and partitioning records, and merging sorted record arrays.
\sys runs on top of Ray, which is implemented in Python and C++.
All of \sys's source code is available at \url{https://github.com/exoshuffle/cloudsort}.

\section{Evaluation}
\subsection{Environment Setup}
\label{sec:eval:cloud-setup}

We run \sys on AWS on a compute cluster configured as follows:
\begin{itemize}
\item $1\times$ \textsf{r6i.2xlarge} master node. This node runs on 8 cores of an Intel Xeon 8375C CPU at 2.9\,GHz, and 64\,GiB memory.
\item $40\times$ \textsf{i4i.4xlarge} worker nodes. Each node runs on 16 cores of an Intel Xeon 8375C CPU at 2.9\,GHz, and 128\,GiB memory. Each node has a directly-attached 3.75\,TB AWS Nitro NVMe SSD.
\item Each node is attached with a 40\,GiB Amazon EBS General Purpose SSD (\textsf{gp3}) volume.
% Cite https://web.archive.org/web/20221008095605/https://docs.aws.amazon.com/AWSEC2/latest/UserGuide/storage-optimized-instances.html
\end{itemize}

The software stack is configured as follows:
\begin{itemize}
\item \textsf{Ubuntu 22.04.1 LTS}, Linux kernel version \textsf{5.15.0-1022-aws}.
\item XFS \textsf{5.13.0} filesystem.
\item Intel oneAPI DPC++/C++ Compiler \textsf{2022.2.0.20220730}.
\item Python \textsf{3.9.13}.
\item Ray \textsf{2.1.0}.
\end{itemize}

We measure the raw system I/O performance on the worker nodes using standard benchmarking tools:
\begin{itemize}
\item Network bandwidth: 25\,Gbps between nodes, benchmarked with \textsf{iperf}.
\item SSD: 2.9\,GB/s read, 2.2\,GB/s write, benchmarked with \textsf{fio}.
% sudo fio --directory=/mnt/data0 --ioengine=psync --name fio_test_file --direct=1 --rw=randwrite --bs=64k --size=1G --numjobs=16 --time_based --runtime=30 --group_reporting --norandommap
% sudo fio --directory=/mnt/data0 --name fio_test_file --direct=1 --rw=randread --bs=64k --size=1G --numjobs=16 --time_based --runtime=30 --group_reporting --norandommap
\end{itemize}

For storage, we use 40 buckets on Amazon S3 and randomly distribute the input and output partitions across the buckets.

% Theoretical time: \theoreticaljct{} seconds.

\subsection{Benchmark Setup}
\label{sec:eval:bench-setup}

\paragraph{Generating Input}
We use \code{gensort} version 1.5 as provided by the Sort Benchmark committee~\cite{gensort}.
We run the command \lstinline|gensort -c -b{offset} {size} {path}| to generate each partition.
\lstinline|{size}| is fixed at $P=\numprint{20000000}$ such that each partition is exactly 2\,GB.
\lstinline|{offset}| takes the values $\{ i\cdot P : 0\leq i < M \}$ where the number of input partitions $M=\numprint{\nummappers{}}$.
\lstinline|{path}| is a unique path in tmpfs.
\lstinline|-c| provides data checksum for validation.
After generating an input file, we randomly choose a bucket and upload the partition to S3.
We use Ray to schedule the \numprint{\nummappers{}} input generation tasks to all \numworkers{} worker nodes.
The result is aggregated as an input manifest file, saved for use by \sys to locate the sort input.

\paragraph{Validating Output}
\sys produces an output manifest file containing the bucket and keys of each output partition on S3.
In each validation task, we first download the output partition to tmpfs, then run the command \lstinline|valsort -o {sumpath} {path}| to validate the ordering of records in each partition.
We use Ray to schedule the \numprint{\numreducers{}} output validation tasks to all \numworkers{} worker nodes.
We concatenate the contents of the summary files from each validation task, then run \lstinline|valsort -s| to validate the total ordering, and generate the total output checksum.
Finally, we compare the output checksum with the input checksum to verify data integrity.

\subsection{Experimental Results}
\label{sec:eval:results}

\subsubsection{Job Completion Time}
\label{sec:eval:results:time}

On November 10, 2022, we ran the \bench in the AWS US West (Oregon, \textsf{us-west-2}) region with the setup described above.
We first generated the input data on Amazon S3, then ran \sys 3 times, each followed by a validation step.
All 3 runs succeeded with the same output checksum as the input, indicating all bytes are preserved in the sort.
\Cref{tab:jct} reports the job completion times of each run.
The average job completion time is \jctseconds{} seconds, or \jct{} hours.

\begin{table}[ht]
\small
  \begin{tabular}{lllll}
    \toprule
    Run
    &
    Map \& Shuffle Time
    &
    Reduce Time
    &
    Total Job Completion Time
    \\
    \midrule
    \#1 % https://wandb.ai/raysort/raysort/runs/sfycimry
    % &
    % 2:35:51\,am
    &
    3509\,s
    &
    1852\,s
    &
    5361\,s
    \\
    \#2 % https://wandb.ai/raysort/raysort/runs/2kcdmn9l
    % &
    % 4:44:57\,am
    &
    3496\,s
    &
    1852\,s
    &
    5348\,s
    \\
    \#3 % https://wandb.ai/raysort/raysort/runs/3sqnfw7i
    % &
    % 7:35:23\,am
    &
    3520\,s
    &
    1906\,s
    &
    5426\,s
    \\
    \midrule
    Average
    &
    3508\,s
    &
    1870\,s
    &
    \jctseconds{}\,s
    \\
    \bottomrule
\end{tabular}
\caption{Job completion times of \sys on the \bench.}
\label{tab:jct}
\end{table}

\begin{figure}[ht]
  \centering
  \includegraphics[width=\linewidth]{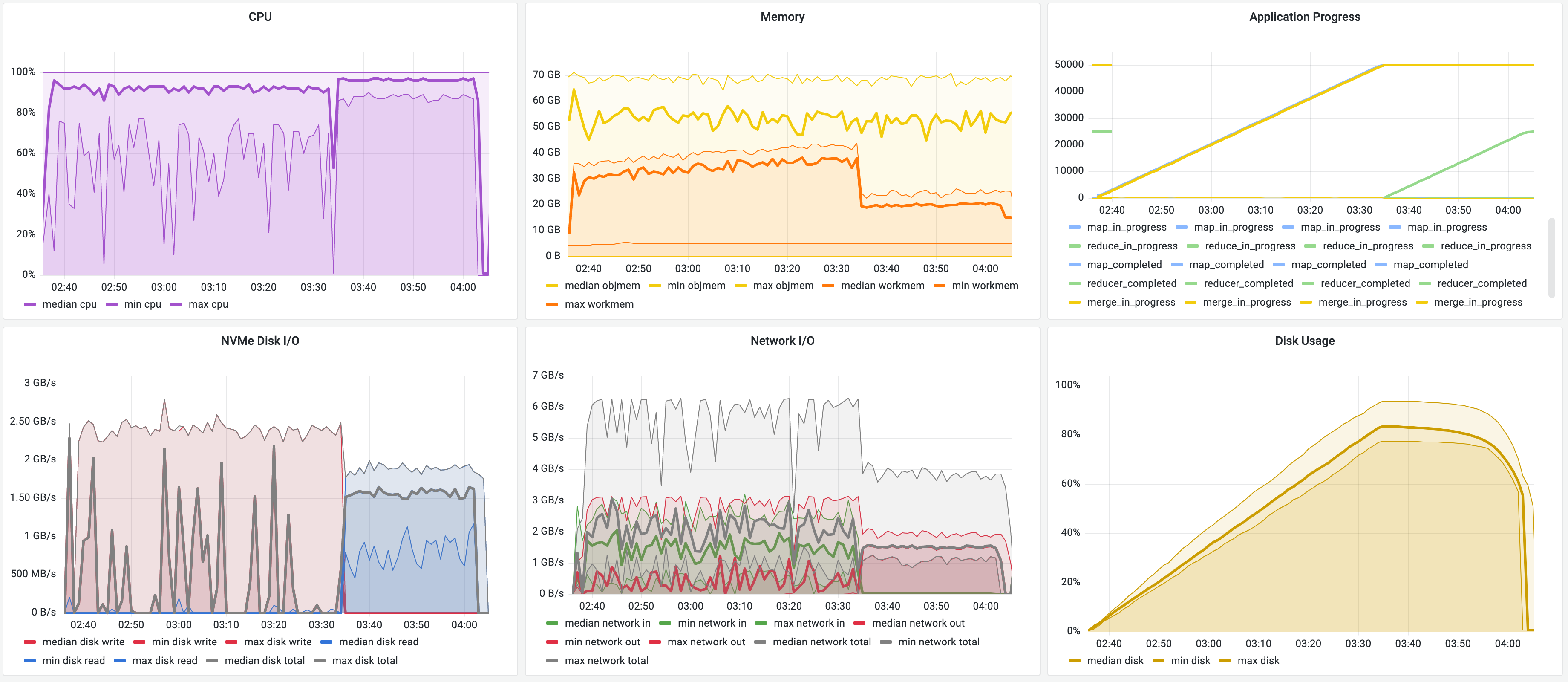}
  \caption{Cluster utilization during run \#1 of the \bench. Each thick line represents the median system utilization of all worker nodes; the highest and lowest lines represent the maximum and minimum utilization among all worker nodes, respectively.}
\label{fig:cluster-utilization}
\end{figure}

\Cref{fig:cluster-utilization} shows the system utilizations of all worker nodes in the compute cluster during run \#1 of the \bench.

% https://raysort.grafana.net/dashboard/snapshot/T23DfAI3HQjb5tOUSB3BGhJJvduFYCDE?orgId=1&from=1668076551000&to=1668081912000

\subsubsection{Total Cost of Ownership}
\label{sec:eval:results:cost}

The total job cost comprises of two parts: compute cost (Amazon EC2), and the storage cost (Amazon S3). The storage cost is further divided into data storage cost and data access cost.

\paragraph{Compute Cost}
The compute cost is calculated as the compute cluster's hourly cost times the job completion time. The total hourly cost is calculated as follows:
\begin{equation}
\label{eq:vmtotalhourlycost}
\begin{aligned}
\text{Total Hourly Compute Cost} &= \text{Master Node Hourly Cost} \\
&+ \text{Worker Node Hourly Cost} \times \text{Number of Workers} \\
&+ \text{EBS Volume Hourly Cost} \times (\text{Number of Workers} + 1)
\end{aligned}
\end{equation}

We obtain the compute instance hourly costs from the Amazon EC2 on-demand pricing information~\cite{aws-ec2-pricing}.
For EBS, we use the Amazon EBS monthly price~\cite{aws-ebs-pricing} divided by the average number of hours in a month ($\frac{365\times 24}{12} = \hoursinmonth$) as the hourly price.
The hourly cost of a 40\,GiB \textsf{gp3} volume is $\$0.08 / \hoursinmonth \times 40 = \$\ebsvolumehourlycost$.
Now we plug the cost variables into \Cref{eq:vmtotalhourlycost}:

\begin{itemize}
\item Master node (\textsf{r6i.2xlarge}) hourly cost is \$\masternodehourlycost{}.
\item Worker node (\textsf{i4i.4xlarge}) hourly cost is \$\workernodehourlycost{}.
\item Number of workers is \numworkers{}.
\item EBS volume hourly cost is \$\ebsvolumehourlycost{}.
\end{itemize}

Hence, the total hourly compute cost is \$\vmtotalhourlycost{}.
We multiply this hourly cost by the job completion time of \jct{} hours to obtain the total compute cost of \$\computetotalcost{}.

\paragraph{Data Storage Cost}
The storage cost comprises of data storage cost and data access cost.
We first consider the data storage cost.
Amazon S3 employs a pay-as-you-go pricing model, i.e. the user does not need to provision storage capacity ahead of time, and only pays for the storage cost of objects based on their sizes and storage duration.
Amazon S3 charges \$0.023 per GB-month for the first 50\,TB, then \$0.022 per GB-month for the next 450\,TB~\cite{aws-s3-pricing}.
Since the total data size is 100\,TB, we take the average price between the first two tiers, i.e. \$\sssgbmonthlycost{} per GB-month, or \$\ssstotalhourlycost{} per hour per 100\,TB.

\begin{itemize}
\item Input: The storage cost of the 100\,TB input data is simply the cost to store 100\,TB for the duration of the sort: $\$\ssstotalhourlycost{} \times \jct{} = \$\datastorageinputcost{}$. 

\item Output: The 100\,TB output data is uploaded to and stored on Amazon S3 during the reduce stage of the sort. We use the duration of the reduce stage as the storage time of the 100\,TB output data. This is an over-estimation because the output partitions are uploaded as the reduce stage progresses, and therefore most of the 100\,TB is stored on S3 for less time than the entire reduce stage duration. \Cref{tab:jct} shows the average reduce stage time is \reduceseconds{} seconds, or \reducetime{} hours. Hence we get the output storage cost: $\$\ssstotalhourlycost{} \times \reducetime{} = \$\datastorageoutputcost{}$. 
\end{itemize}

Adding up the input and output data storage cost, we get the total data storage cost: \$\datastoragetotalcost{}.

\paragraph{Data Access Cost}
We consider GET and PUT requests to Amazon S3.
\sys downloads the 100\,TB input data in \numprint{\nummappers} map tasks. Each map task downloads a 2\,GB input partition in 16\,MiB chunks, resulting in \sssnumgetpertask{} GET requests per task, or \numprint{\ssstotalget} GET requests in total.
Amazon S3 charges \$0.0004 per 1000 GET requests~\cite{aws-s3-pricing}.
Hence the total GET cost is \$\sssgettotalcost{}.

\sys uploads the output data in \numprint{\numreducers} reduce tasks.
Each reduce task uploads approximately 4\,GB data in 100\,MB chunks, resulting in \sssnumputpertask{} PUT requests, or \numprint{\ssstotalput} PUT requests in total.
Amazon S3 charges \$0.005 per 1000 PUT requests~\cite{aws-s3-pricing}.
Hence the total PUT cost is \$\sssputtotalcost{}.

The actual number of requests could be marginally higher due to request failures and retries, but the amount should be negligible.
Hence, the total data access cost is \$\dataaccesstotalcost{}.

\paragraph{Total Cost of Ownership}
Adding up the compute cost and storage cost, we get the total cost of ownership for the \bench: \$\totalcost{}. \Cref{tab:totalcost} presents a summary of the cost analysis.

\begin{table}[ht]
\small
  \begin{tabular}{llll}
    \toprule
    Service
    &
    Unit Price
    &
    Amount
    &
    Total Price
    \\
    \midrule
    Compute VM Cluster
    &
    \$\vmtotalhourlycost{} / hr
    &
    \jct{} hours
    &
    \$\computetotalcost{}
    \\
    \midrule
    Data Storage (Input)
    &
    \$\ssstotalhourlycost{} / hr
    &
    \jct{} hours
    &
    \$\datastorageinputcost{}
    \\
    Data Storage (Output)
    &
    \$\ssstotalhourlycost{} / hr
    &
    \reducetime{} hours
    &
    \$\datastorageoutputcost{}
    \\
    Data Access (Input)
    &
    \$0.0004 / 1000 requests
    &
    \numprint{\ssstotalget} requests
    &
    \$\sssgettotalcost{}
    \\
    Data Access (Output)
    &
    \$0.005 / 1000 requests
    &
    \numprint{\ssstotalput} requests
    &
    \$\sssputtotalcost{}
    \\
    \midrule
    Total
    &
    --
    &
    --
    &
    \$\totalcost{}
    \\
    \bottomrule
\end{tabular}
\caption{Cost breakdown of \sys on the \bench.}
\label{tab:totalcost}
\end{table}

\begin{acks}
This work is done in the Sky Computing Lab at UC Berkeley, sponsored by Astronomer, Google, IBM, Intel, Lacework, Nexla, Samsung SDS, and VMware.
This work is done in collaboration with Anyscale.
\end{acks}

\bibliographystyle{ACM-Reference-Format}
\bibliography{_main}

\end{document}